*Article*

# High-pressure synthesis and superconducting properties of NaCl-type In$_{1-x}$Pb$_x$Te ($x$ = 0–0.8)

**Masayoshi Katsuno[1], Rajveer Jha[1], Kazuhisa Hoshi[1], Ryota Sogabe[1], Yosuke Goto[1] and Yoshikazu Mizuguchi [1,\*]**

[1] Department of Physics, Tokyo Metropolitan University
\* Correspondence: mizugu@tmu.ac.jp

**Abstract:** We have investigated the Pb-substitution effect on the superconductivity of NaCl-type In$_{1-x}$Pb$_x$Te samples obtained using high-pressure synthesis. Polycrystalline samples with $x$ = 0–0.8 were obtained, and the lattice parameter systematically increased by Pb substitution. The superconducting transition temperature ($T_c$ ~3 K for $x$ = 0) increased to 5.3 K (for $x$ = 0.4) by Pb substitutions. According to the comparison of the electronic states between NaCl-type InTe and PbTe, we consider that the Pb substitution increases electron carriers. Possible mechanisms of the enhancement of $T_c$ in this system have been discussed on the basis of conventional pairing mechanism and the possibility of the relation to the valence-skipping state of In.

## 1. Introduction

Metal chalcogenides have drawn great attention in the research community of superconductivity because of the observation of high transition temperature ($T_c$) and unconventional pairing mechanisms in those materials [1–14]. Particularly, SnTe has been extensively studied because it is a topological crystalline insulator [15] and shows superconductivity when the Sn site was substituted by In or Ag [8–14]. Although non-doped SnTe also shows superconductivity below 0.3 K [16], hole doping effectively increases $T_c$ to 4.8 K in the In-substituted system [8–12] and to 2.4 K in the Ag-substituted system [13,14]. In addition, possible topological superconductivity has been proposed for the Sn$_{1-x}$In$_x$Te system [15,17]. Due to these findings, Sn$_{1-x}$In$_x$Te has been one of hot-topic materials in the superconductivity community. Recently, superconductivity in the In-rich phases of Sn$_{1-x}$In$_x$Te has been reported [18,19]. The solubility limit of In for the Sn site can be extended to $x$ = 1.0 using high-pressure synthesis [18,19]. Interestingly, the sample with $x$ = 1.0 (pure InTe) with a NaCl-type structure also shows superconductivity ($T_c$ ~3 K). As shown in Fig. 1(a,b), InTe exhibits a pressure-induced structural transition from a TlSe-type to NaCl-type structure. Although In$^+$ and In$^{3+}$ occupy different sites in the TlSe structure, those In occupies single site in the NaCl structure. Namely, the valence state of In in NaCl-type InTe can be regarded as the valence-skipping state of In$^+$ and In$^{3+}$ [19]. Hence, NaCl-type InTe is a good material for investigating the friendship between valence skipping and superconductivity. Having considered the inducement of superconductivity in a valence skipping material BaBiO$_3$ by carrier doping (or introduction of site disorder) by Pb or K substitutions [20,21], we can expect an increase in $T_c$ by element substitution to InTe. As a fact, in Ref. 18, an increase in $T_c$ (to ~5 K) by Se substitution in InTe$_{1-x}$Se$_x$ was reported.

In this study, we have investigated the effect of Pb substitution on the superconductivity of In$_{1-x}$Pb$_x$Te. PbTe ($x$ = 1.0) has a NaCl-type structure and is a band insulator. PbTe is also a good material to discuss the relationship between valence skipping states and superconductivity. Although PbTe itself is not a valence skipping material, Tl-doped system (Pb$_{1-x}$Tl$_x$Te) shows a correlation between valence skipping states of Tl and superconducting properties [22]. On the basis of these notable characteristics of InTe and PbTe, we have decided to study their solution system In$_{1-x}$Pb$_x$Te. We observed an increase in $T_c$ by Pb substitution and established a dome-shaped superconductivity phase diagram. The possible mechanisms for the increase in $T_c$ by Pb substitution are discussed.

## 2. Results

Powder X-ray diffraction (XRD) patterns for $In_{1-x}Pb_xTe$ are shown in Fig. 1(c). The XRD peak position shifts toward lower angles by Pb substitution, which indicates a lattice expansion by Pb substitution. The XRD patterns were refined using a cubic NaCl-type model ($Fm$-$3m$, #225, $O_h^5$) by Rietveld refinement with the RIETAN-FP software [23]. Lattice parameters obtained from refinements are plotted in Fig. 1(d) as a function of Pb concentration $x$. The lattice parameter monotonically increases with increasing Pb concentration.

Figure 2(a) shows the temperature dependences of magnetic susceptibility for $x$ = 0–0.8 measured after both ZFC (zero-field cooling) and FC (field cooling). Note that PbTe ($x$ = 1.0) is an insulator and does not show a superconducting transition. For all the samples, superconducting (diamagnetic) signals are observed, and a large shielding volume fraction is estimated, which indicates the emergence of bulk superconductivity in $x$ = 0–0.8. From the temperature dependences of magnetic susceptibility, a $T_c$ was estimated as the temperature where susceptibility signal begins to decrease, and an irreversibility temperature ($T_{irr}$) was estimated as the temperature where the ZFC and FC curves are clearly separated. In bulk samples, $T_{irr}$ generally corresponds to the temperature at which zero resistivity states appear. The estimated $T_c$ and $T_{irr}$ are plotted in Fig. 2(b) as a function of $x$. A dome-shaped phase diagram on superconductivity was obtained. The highest $T_c$ of 5.3 K is obtained for $x$ = 0.4, but those for $x$ = 0.5–0.8 are comparable to that for $x$ = 0.4. Notably, the difference between $T_c$ and $T_{irr}$ is very small for $x$ = 0 and 0.4−0.8. This indicates that the evolution of superconductivity states is homogeneous in the examined samples.

Figure 3(a) shows the temperature dependences of electrical resistivity for $x$ = 0–1.0. PbTe ($x$ = 1.0) exhibits an insulating behavior. For $x$ = 0.8, the insulating behavior of pure PbTe is suppressed, and the temperature dependence becomes almost temperature-independent. With increasing In concentration (with decreasing $x$), electrical resistivity decreases and metallic conductivity is induced. We notice that the temperature dependence of electrical resistivity for PbTe shows a flat dependence at low temperatures, which has also been reported in a previous study [24]. The zoomed figure for low-temperature data is displayed in Fig. 3(b). Zero-resistivity states are observed for all the superconducting samples.

To investigate the nature of electronic states, we measured room-temperature Seebeck coefficient for all the samples. In Fig. 4, Pb concentration ($x$) dependence of the Seebeck coefficient is plotted. Since PbTe is a band insulator, whose electronic structure is shown in Fig. 6, large negative Seebeck coefficient of -342 μV/K is observed. By In substitution, Seebeck coefficient becomes very small and positive in sign even for $x$ = 0.8. This suggests that only 20% substitution of In for the Pb site can generate large amount of hole carriers. This is consistent with what is expected from band structure shown in Fig. 6. With increasing In concentration (with decreasing $x$), the absolute value of Seebeck coefficient decreases and increases at $x$ = 0.

To investigate the superconducting properties of $In_{1-x}Pb_xTe$ under magnetic field, electrical resistivity was measured under magnetic fields. Figure 5(a) shows the temperature dependences of electrical resistivity for $x$ = 0.4, in which the highest $T_c^{onset}$ was observed in susceptibility measurements, under magnetic fields up to 3.0 T. To obtain an upper critical field ($\mu_0H_{c2}$) phase diagram, $T_c$ was estimated as the temperature where the resistivity becomes 80% of normal state resistivity just above $T_c^{onset}$. The $H_{c2}$-temperature data were analyzed by the Werthamer-Helfand-Hohemberg (WHH) model [25] as shown in Fig. 5(b), and the estimated $\mu_0H_{c2}$ ($T$ = 0 K) was 3.1 T.

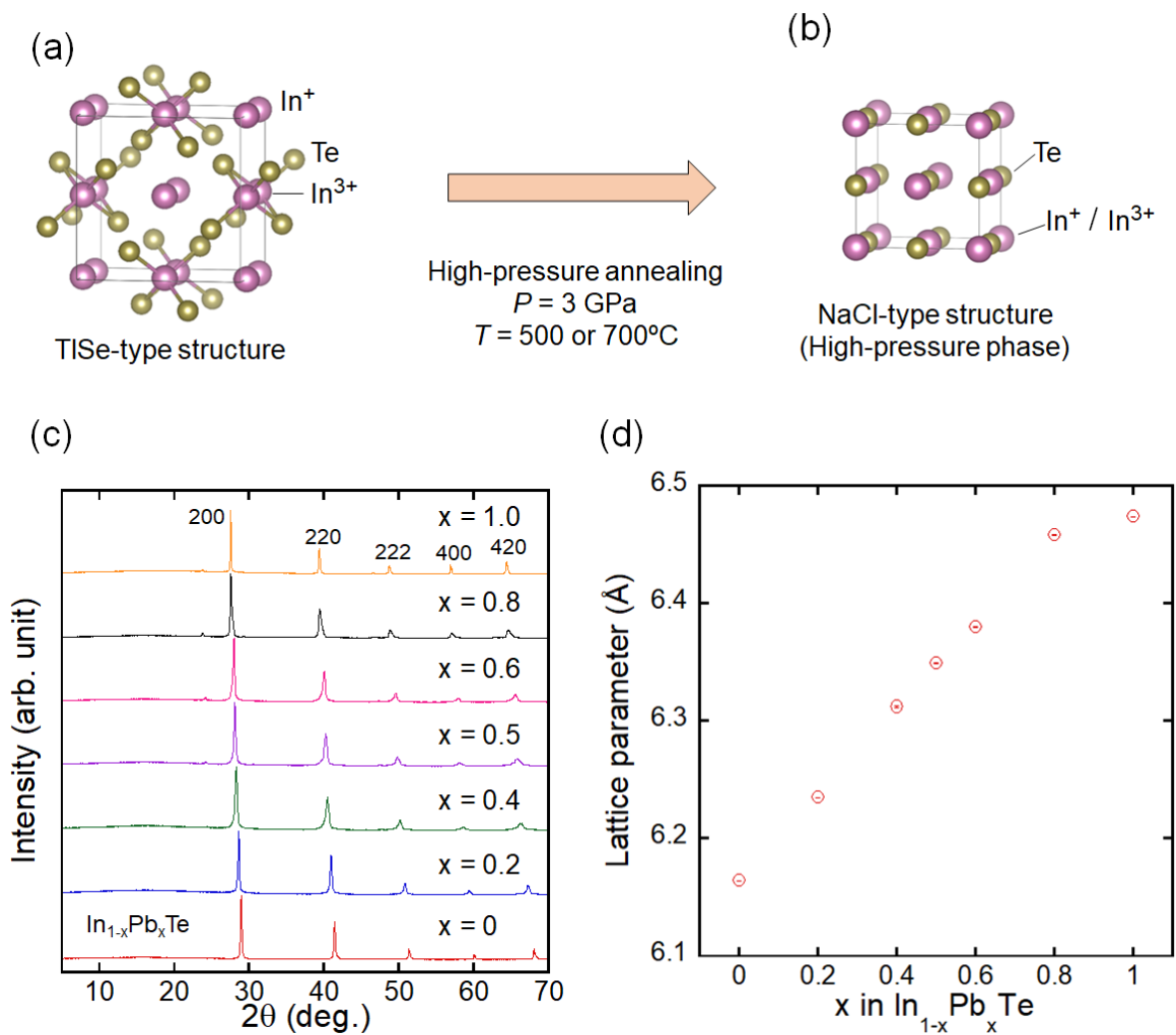

Figure 1. (a,b) Schematic images of crystal structure of InTe: (a) ambient-pressure phase with a TlSe-type structure and (b) high-pressure phase with a NaCl-type structure. (c) Powder X-ray diffraction patterns for In$_{1-x}$Pb$_x$Te. Numbers in this figure are Miller indices. (d) Pb concentration ($x$) dependence of lattice parameter of a for In$_{1-x}$Pb$_x$Te.

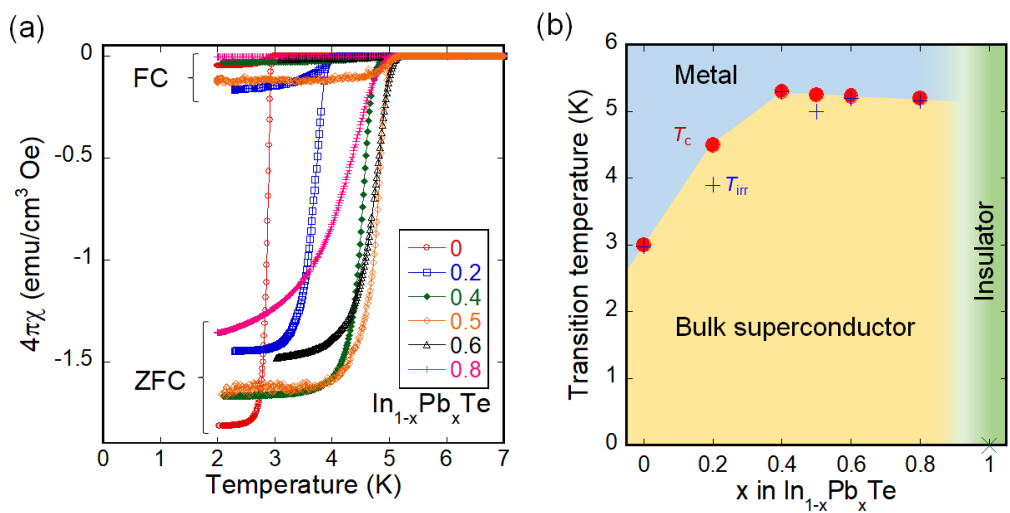

Figure 2. (a) Temperature dependences of magnetic susceptibility $4\pi\chi$ for In$_{1-x}$Pb$_x$Te. (b) Phase diagram of In$_{1-x}$Pb$_x$Te with a NaCl-type structure. $T_c$ and $T_{irr}$ were determined from the magnetic susceptibility data. The cross symbol in the figure ($x$ = 1.0) indicates a non-superconducting phase.

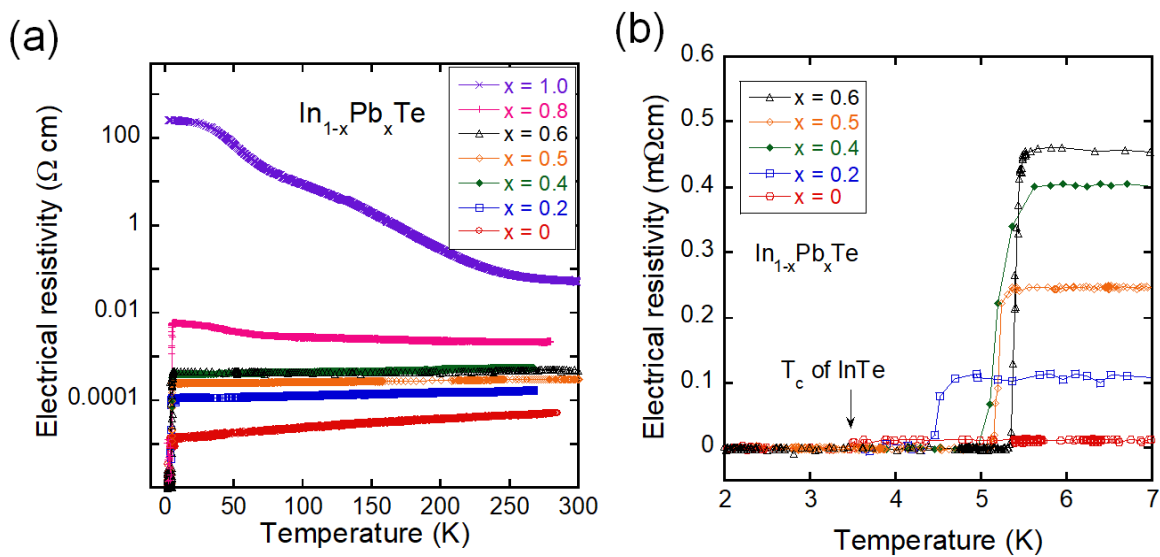

Figure 3. (a) Temperature dependences of electrical resistivity for In$_{1-x}$Pb$_x$Te. (b) Low-temperature data of the temperature dependences of electrical resistivity for In$_{1-x}$Pb$_x$Te.

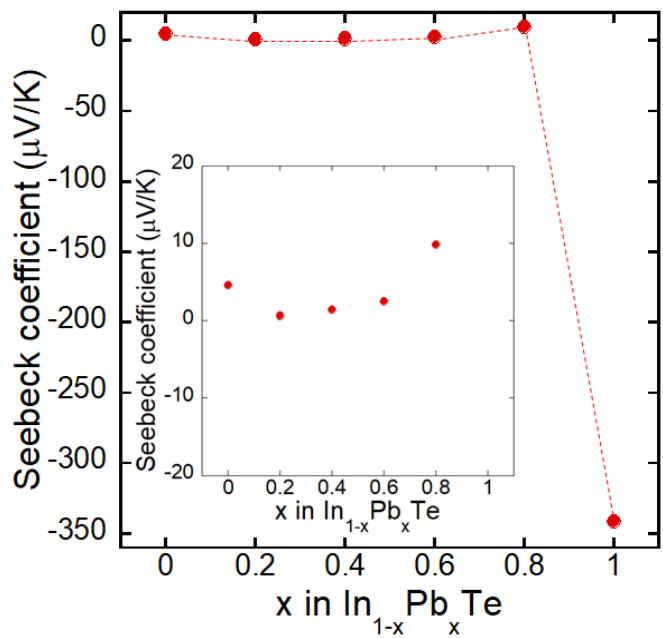

Figure 4. Pb concentration ($x$) dependence of Seebeck coefficient for In$_{1-x}$Pb$_x$Te. The inset shows a zoomed plot for $x$ = 0–0.8.

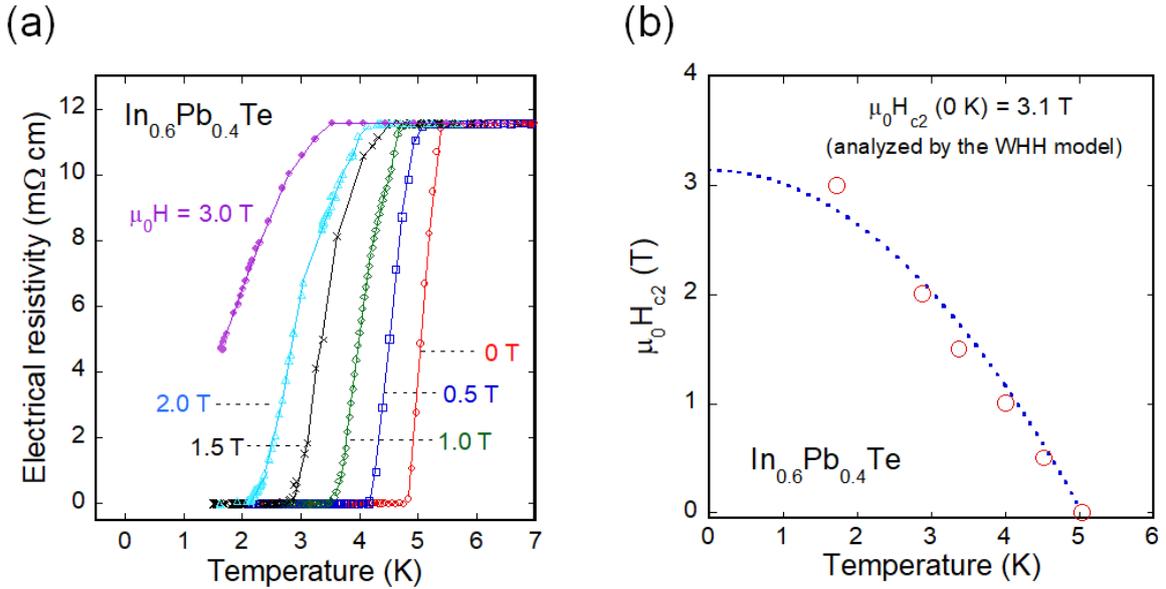

Figure 5. (a) Temperature dependences of electrical resistivity for $x = 0.4$ under magnetic fields of 0–3 T. (b) Temperature dependence of $\mu_0 H_{c2}$ for $x = 0.4$ where $T_c$ was estimated as the temperature where the resistivity becomes 80% of normal state resistivity just above $T_c^{onset}$. The dashed line is the fitting result by the WHH model.

## 3. Discussion

In this article, we reported that $T_c$ of NaCl-type InTe ($T_c \sim 3$ K) can be increased to 5.3 K by Pb substitution for the In site. Here, we discuss about this increase in $T_c$ with two possible scenarios.

The first scenario is an explanation based on density of states. In Fig. 6, total density of states (DOS) taken from CompES-X, MatNavi, which provides band structure calculated with a typical structural parameter of the materials by first-principle calculations with the VASP program. Assuming conventional electron-phonon mechanism [26], an increase in DOS at Fermi energy ($E_F$) basically increases $T_c$. Let us consider the emergence of superconductivity from the right side of the phase diagram (from PbTe, $x = 1.0$). Since PbTe is a band insulator, $E_F$ locates in a band gap. However, there is a very high DOS in the valence band top. The high DOS may be working positively to the superconductivity of $In_{1-x}Pb_xTe$. Actually, the evolution of Seebeck coefficient in $In_{1-x}Pb_xTe$ suggests that large amount of hole carriers was generated for all the doped samples ($x = 0$-0.8) on the basis of the relationship between carrier concentration and Seebeck coefficient ($S$) in metal [27]:

$$S = \frac{8\pi^2 k_B^2 m^* T}{3eh^2}\left(\frac{\pi}{3n}\right)^{\frac{2}{3}},$$

where $k_B$ is the Boltzmann constant, $m^*$ is the effective mass of the carrier, $e$ is the elementary charge, $h$ is the Planck constant, and $n$ is the carrier concentration. We can see that the evolution of Seebeck coefficient by changing Pb concentration (Fig. 4) shows the correlation with the $T_c$-$x$ phase diagram. For $x = 0$, the Seebeck coefficient is slightly larger than those for $x = 0.2$–0.8. Actually, the $E_F$ for InTe locates at the left shoulder of the DOS peak. Although the first scenario looks consistent with what we have observed in the present system, the difference in $T_c$ between $x = 0$ and $x = 0.4$–0.8 is large on the basis of small change in DOS at $E_F$, which is expected from the results of Seebeck coefficient. Therefore, we may have to consider the second scenario which is based on the valence-skipping states of In.

Namely, in the second scenario, we assume that the valence-skipping states of In plays an important role in the emergence of superconductivity. As introduced in the introduction part, it has been proposed that the valence-skipping states in $Pb_{1-x}Tl_xTe$ is positively linked to the superconductivity [22]. Although the solution concentration of Tl in $Pb_{1-x}Tl_xTe$ is very limited, our $In_{1-x}Pb_xTe$ system has a perfect solution from $x = 0$ to 1.0. Notably, InTe ($x = 0$) has been proposed as a valence-skipping material [18]. From those facts, we consider that InTe is in the vicinity of charge-ordered phase. In doped $BaBiO_3$ [20,21], charge-density-wave ordering is suppressed by carrier doping, and high-$T_c$ superconductivity is induced. It has been believed that the high-$T_c$ is achieved by the pairing mediated by the negative-$U$ interaction caused by the valence-skipping states of Bi [28]. If the pairing mechanisms of $In_{1-x}Pb_xTe$ is similar to those in the doped $BaBiO_3$, we expect the presence of an ordered (possibly a charge-ordered insulating phase) for the InTe-based system. Since NaCl-type InTe sample can be prepared under high pressure only, strain effect and/or site disorder (partial substitution of In and Te to each site) may exist in the present sample with $x = 0$. These factors may suppress an ordered state, if there was such a charge ordered state. The ordered state can be further suppressed by Pb substitutions, which should increase $T_c$ as observed for $x = 0.4$–$0.8$ in this study. For both scenarios based on conventional and valence-skipping mechanisms, we need further experiments to solve the mechanisms of superconductivity in the $In_{1-x}Pb_xTe$ system.

In conclusion, we have synthesized NaCl-type $In_{1-x}Pb_xTe$ samples using high-pressure synthesis method. From the evolution of lattice parameter, monotonous lattice expansion was confirmed by Pb substitution. For $x = 0$–$0.8$, bulk superconductivity was observed, which was confirmed from magnetic susceptibility and electrical resistivity measurements. $T_c$ was the highest for $x = 0.4$. From the resistivity measurements under magnetic fields, upper critical field $\mu_0 H_{c2}$ ($T = 0$ K) was estimated as 3.1 T. We have proposed two scenarios to explain the increase in $T_c$ by Pb substitution: the first scenario is based on the conventional mechanisms (phonon-mediated BCS), and the second scenario is based on the valence-skipping states of In, which was optimally suppressed by Pb substitution.

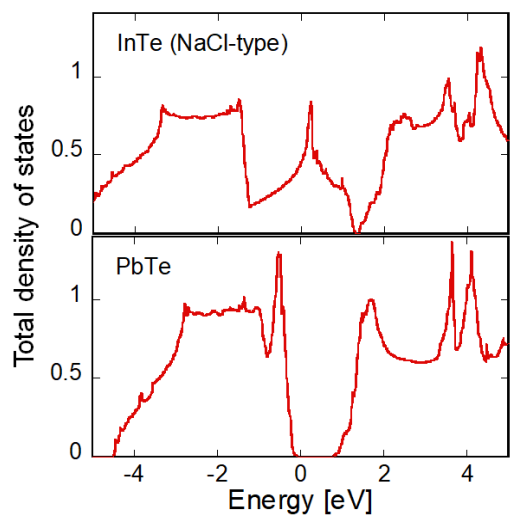

Figure 6. Calculated total density of states for NaCl-type InTe and PbTe. The output data of the density of states from the database CompES-X, MatNavi, https://compes-x.nims.go.jp/index.html, [accessed 2019-02-13]. The calculation conditions are listed below. Lattice parameters: $a = 6.139$ Å for InTe and $a = 6.461$ Å for PbTe. Method: Projector Augmented Wave (PAW). Program: VASP.

## 4. Materials and Methods

Polycrystalline powders of $In_{1-x}Pb_xTe$ were prepared by melting a mixture of In (99.99%), Pb (99.9%), and Te (99.999%) grains with a nominal composition of $In_{1-x}Pb_xTe$. The mixture was sealed in an evacuated quartz tube and heated at 800 °C for 10 h. The obtained sample was ground and pelletized into a pellet with a diameter of 5 mm. The pellet was placed in a high-pressure cell, which is composed of a BN sample capsule, a carbon heater capsule, electrodes, and a pyrophyllite cubic cell. The high-pressure synthesis was performed using 180 ton cubic-anvil-type system (CT factory). The pressure used for the high-pressure synthesis was 3 GPa, and the heating temperature was 700 ºC for $x = 0$ and 500 ºC for $x = 0.2$–$0.8$. The PbTe ($x = 1.0$) sample was prepared by melting in an evacuated quartz tube without carrying out high-pressure synthesis.

Powder X-ray diffraction (XRD) was performed by the $\theta$-$2\theta$ method using MiniFlex600-D/tex-Ultra (RIGAKU) with a CuKα radiation. The crystal structure parameters were refined using the Rietveld method with RIETAN-FP [23]. The crystal structure image was depicted using VESTA [29].

The temperature dependence of magnetic susceptibility was measured using a superconducting quantum interference device (SQUID) magnetometer (MPMS-3, Quantum Design) with a typical applied field of 10 Oe after zero-field cooling (ZFC) and field cooling (FC). Electrical resistivity measurements were performed on cryostat (GM refrigerator) equipped with a superconducting magnet (designed by AXIS/Thermal Block). The temperature dependence of electrical resistivity was measured by a four-terminal method. Au wires with a diameter of 25 μm were attached on the polished surface of the samples by using Ag pastes. A typical current of 1 mA was used for the resistivity measurements. Seebeck coefficient was obtained from the linear slope of the thermo-electromotive force ($\Delta V$) versus the temperature difference ($\Delta T$) plots.

**Author Contributions:** conceptualization, Y.M.; methodology, M.K.; R.J.; K.H.; R.S.; Y.G. and Y.M.; validation, M.K. and Y.M.; formal analysis, Y.M.; investigation, M.K.; R.J.; K.H.; R.S.; Y.G. and Y.M.; resources, Y.G. and Y.M.; data curation, M.K. and Y.M.; writing—original draft preparation, Y.M.; writing—review and editing, Y.M.; visualization, M.K.; R.J. and Y.M.; supervision, Y.G. and Y.M.; project administration, Y.M.; funding acquisition, Y.M.

**Funding:** This work was partly supported by Grants-in-Aid for Scientific Research by JSPS (Nos. 15H05886, 19K15291, and 18KK0076) and Advanced Research Program under the Human Resources Funds of Tokyo.

**Acknowledgments:** The authors thank Hase, I. and Kobayashi, K. for fruitful discussion.

**Conflicts of Interest:** The authors declare no conflict of interest.

**References**

1. Hsu, F. C.; Luo, J. Y.; Yeh, K. W.; Chen, T. K.; Huang, T. W.; Wu, P. M.; Lee, Y. C.; Huang, Y. L.; Chu, Y. Y.; Yan, D. C.; Wu, M. K. Superconductivity in the PbO-type structure α-FeSe. *Proc. Natl. Acad. Sci. U.S.A.* 2008, 105, 14262.
2. Mizuguchi, Y.; Takano, Y. Review of Fe chalcogenides as the simplest Fe-based superconductor. *J. Phys. Soc. Jpn.* 2010, 79, 102001.
3. Hor, Y. S.; Williams, A. J.; Checkelsky, J. G.; Roushan, P.; Seo, J.; Xu, Q.; Zandbergen, H. W.; Yazdani, A.; Ong, N. P.; Cava, R. J. Superconductivity in $Cu_xBi_2Se_3$ and its Implications for Pairing in the Undoped Topological Insulator. *Phys. Rev. Lett.* 2010, 104, 057001.
4. Sasaki, S.; Kriener, M.; Segawa, K.; Yada, K.; Tanaka, Y.; Sato, M.; Ando, Y. Topological Superconductivity in $Cu_xBi_2Se_3$. *Phys. Rev. Lett.* 2011, 107, 217001.
5. Mizuguchi, Y. Review of superconductivity in $BiS_2$-based layered materials. *J. Phys. Chem. Solids* 2015, 84, 34.
6. Mizuguchi, Y. Material Development and Physical Properties of $BiS_2$-Based Layered Compounds. *J. Phys. Soc. Jpn.* 2019, 88, 041001.
7. Ren, Z.; Kriener, M.; Taskin, A. A.; Sasaki, S.; Segawa, K.; Ando, Y. Anomalous metallic state above the upper critical field of the conventional three-dimensional superconductor $AgSnSe_2$ with strong intrinsic disorder. *Phys. Rev. B* 2013, 87, 064512.
8. Erickson, A. S.; Chu, J. H.; Toney, M. F.; Geballe, T. H.; Fisher, I. R. Enhanced superconducting pairing interaction in indium-doped tin telluride. *Phys. Rev. B* 2009, 79, 024520.
9. Balakrishnan, G.; Bawden, L.; Cavendish, S.; Lees, M. R. Superconducting properties of the In-substituted topological crystalline insulator SnTe. *Phys. Rev. B* 2013, 87, 140507.
10. Novak, M.; Sasaki, S.; Kriener, M.; Segawa, K.; Ando, Y. Unusual nature of fully gapped superconductivity in In-doped SnTe. Phys. Rev. B 2013, 88, 140502.
11. Zhong, R. D.; Schneeloch, Shi, X. Y.; Xu, Z. J.; Zhang, C.; Tranquada, J. M.; Li, Q.; Gu, G. D. Optimizing the superconducting transition temperature and upper critical field of $Sn_{1-x}In_xTe$. *Phys. Rev. B* 2013, 88, 020505.

12. Haldolaarachchige, N.; Gibson, Q.; Xie, W.; Nielsen, M. B.; Kushwaha, S.; Cava, R. J. Anomalous composition dependence of the superconductivity in In-doped SnTe. *Phys. Rev. B* 2016, 93, 024520.
13. Mizuguchi, Y.; Miura, O. High-Pressure Synthesis and Superconductivity of Ag-Doped Topological Crystalline Insulator SnTe ($Sn_{1-x}Ag_xTe$ with $x$ = 0–0.5). *J. Phys. Soc. Jpn.* 2016, 85, 053702.
14. Mizuguchi, Y.; Yamada, A.; Higashinaka, R.; Matsuda, T. D.; Aoki, Y.; Miura, O.; Nagao, M. Specific Heat and Electrical Transport Properties of $Sn_{0.8}Ag_{0.2}Te$ Superconductor. *J. Phys. Soc. Jpn.* 2016, 85, 103701.
15. Ando, Y.; Fu, L. Topological Crystalline Insulators and Topological Superconductors: From Concepts to Materials. *Annu. Rev. Condens. Matter Phys.* 2015, 6, 361.
16. Mathur, M. P.; Deis, D. W.; Jones, C. K.; Carr Jr., W. J. Superconductivity as a function of carrier density and magnetic spin concentration in the SnTe-MnTe system. *J. Phys. Chem. Solids* 1973, 34, 183.
17. Sato, T.; Tanaka, Y.; Nakayama, K.; Souma, S.; Takahashi, T.; Sasaki, S.; Ren, Z.; Taskin, A. A.; Segawa, K.; Ando, Y. Fermiology of the Strongly Spin-Orbit Coupled Superconductor $Sn_{1-x}In_xTe$: Implications for Topological Superconductivity. *Phys. Rev. Lett.* 2013, 110, 206804.
18. Kobayashi, K.; Ai, Y.; Jeschke, H. O.; Akimitsu, J. Enhanced superconducting transition temperatures in the rocksalt-type superconductors $In_{1-x}Sn_xTe$ (x ≤ 0.5). *Phys. Rev. B* 2018, 97, 104511.
19. Kriener, M.; Kamitani, M.; Koretsune, T.; Arita, R.; Taguchi, Y.; Tokura, Y. Tailoring band structure and band filling in a simple cubic (IV, III)-VI superconductor. *Phys. Rev. Materials* 2018, 2, 044802.
20. Sleight, A. W.; Gillson, J. L.; Bierstedt, P. E. High-temperature superconductivity in the $BaPb_{1-x}Bi_xO_3$ systems. *Solid State Commun.* 1975, 17, 27.
21. Cava, R. J.; Batlogg, B.; Krajewski, J. J.; Farrow, R.; Rupp Jr, L. W.; White, A. E.; Short, K.; Peck, W. F.; Kometani, T. Superconductivity near 30 K without copper: the $Ba_{0.6}K_{0.4}BiO_3$ perovskite. *Nature* 1988, 332, 814.
22. Matsushita, Y.; Bluhm, H.; Geballe, T. H.; Fisher, I. R. Evidence for Charge Kondo Effect in Superconducting Tl-Doped PbTe. *Phys. Rev. Lett.* 2005, 94, 157002.
23. Izumi, F.; Momma, K. Three-Dimensional Visualization in Powder Diffraction. Solid State Phenom. 2007, 130, 15.
24. Martin, J.; Nolas, G. S.; Zhang, W.; Chen, L. PbTe nanocomposites synthesized from PbTe nanocrystals. Appl. Phys. Lett. 2007, 90, 222112.
25. Werthamer, N. R.; Helfand, E.; Hohemberg, P. C. Temperature and Purity Dependence of the Superconducting Critical Field, $H_{c2}$. III. Electron Spin and Spin-Orbit Effects. *Phys. Rev.* 1966, 147, 295.
26. Bardeen, J.; Cooper, L.N.; Schrieffer, J.R. Theory of superconductivity. Phys. Rev. 1957, 108, 1175.
27. Snyder, G. J.; Toberer, E. S. Complex thermoelectric materials. *Nat. Mater.* 2008, 7, 105.
28. Hase, I.; Yanagisawa, T. Madelung energy of the valence-skipping compound $BaBiO_3$. Phys. Rev. B 2007, 76, 174103.
29. Momma, K.; Izumi, F. VESTA: a three-dimensional visualization system for electronic and structural analysis. *J. Appl. Crystallogr.* 2008, 41, 653.